\documentclass[]{tMOP2e}
\begin{document}

\title{Coherent backscattering under condition of electromagnetically induced transparency}
\author{I.M. Sokolov${}^{1,2}$, D.V. Kupriyanov${}^{1}$ and M.D. Havey${}^{3}$\\
\small $^{1}$Department of Theoretical Physics, State Polytechnic
University, 195251, St.-Petersburg, Russia \\ \small $^{2}$Institute for
Analytical Instrumentation, RAS, 198103, St.-Petersburg, Russia  \\\small$^{3}$Department of Physics, Old Dominion University,
Norfolk, VA 23529}
%\affiliation

\date{\today }
\maketitle

\begin{abstract}
We consider the influence of a resonant control field on weak localization of light in ultracold atomic ensembles. Both steady-state and pulsed light excitation are considered. We show that the presence of a control field essentially changes the type of interference effects which occur under conditions of multiple scattering.  For example, for some scattering polarization channels the presence of a control field can cause destructive interference through which the enhancement factor, normally considered to be greater than one, becomes less than one.

\end{abstract}

\section{Introduction}

Interference effects under multiple light scattering in optically thick disordered media
have been subject to intense investigation for almost three decades.
Coherent backscattering (CBS), which is closely related to weak localization, is one of the striking
examples of the types of effects which can occur. CBS manifest itself in an enhancement in the
intensity of light scattered in the nearly backwards direction. Wave
scattering in this direction along reciprocal, or time-reversed, multiple
scattering paths preserves the relative phase, which results in constructive
interference. The first detailed observations and analysis of this effect for light were
made in \cite{1}-\cite{3} and by now CBS in solids and liquids has
been investigated in detail \cite{4}-\cite{8}.

Observation of CBS of light in atomic gases is complicated because atomic motion
causes random phase shifts of scattering waves which are different for
reciprocal paths. For this reason, the weak localization has been observed only for a cold
atomic ensemble. The first experiments on CBS in ultracold atomic samples
prepared in a magneto-optical trap \cite{9}-\cite{11} shown that there are a
range of interesting features of this phenomena compared with CBS observed earlier
for solids and liquids \cite{4}-\cite{8}. These features, which have their origin in the atomic nature of the scatterers,  could not be understood and quantitatively described by approaches developed previously
for classical scatterers such as powders and suspensions.

Detailed treatments of CBS, taking into account the quantum nature of atomic
scatterers, was developed by several groups \cite{12}-\cite{19.0}.
In these papers different aspects of weak localization were considered.
Particularly it was shown that one can have a strong influence on the observed interferences by
reinforcing those scattering channels which lead to interference and suppressing
those which do not. Such desirable effects can be realized for example by
polarization of atomic angular moments. By means of optical orientation
effects it is possible to collect all atoms in one Zeeman sublevels
ensuring the fulfillment of optimal interference. Similar effects can be
achieved by applying an auxiliary static magnetic field and tuning the frequency of
the light in such a way that light would interact only with the desire Zeeman
sublevel \cite{19}.

Experiment \cite{19} has shown that a static external field
influences the process of multiple scattering and the associated interferences. At the same time it is well known that control of optical properties of matter by means of auxiliary quasiresonant electromagnetic fields is
much more effective than by a static one. Electromagnetically induced atomic
coherence changes the optical properties of atomic samples in sometimes
dramatic ways, and is responsible for such effects as population trapping,
electromagnetically induced transparency (EIT), "slow light"\,, "stopped
light," to name a few \cite{20}-\cite{22}.

In this paper we are going to analyze how a resonant control field influences
coherent backscattering of light.  In particular, we consider CBS effects under conditions of electro-magnetically
induced transparency. Our attention is focused on the spectral
dependence of the relative amplitude of the backscattering cone in a steady-state
regime, i.e. on the dependence of dthe enhancement factor on frequency of scattered
light which assumed to be monochromatic. We show that a control field can
essentially change the type of interference and even can cause destructive
interference. For some scattering polarization channels and for some
detunings of the probe light, the enhancement factor can be less than one. With
the developed knowledge of the spectrum we also consider the dynamics of CBS in the case of
pulsed probe radiation.

\section{Basic assumptions}

One of main quantitative characteristics of CBS is the enhancement factor, which
determines the relative contribution of interference effects to the total
scattered light intensity. In experiment it is measured as the relative amount of light
intensity scattered into a given direction inside CBS cone to the background
intensity registered outside the cone. In theory it is more convenient to
evaluate the differential cross section of scattering from the input to outgoing
mode and calculate the enhancement factor as the ratio of this cross section to
its non-interfering part.

Our theoretical approach allowing calculation of the differential cross section of
light scattered from optically thick ultracold atomic ensembles is described
in details in a series of papers \cite{13}, \cite{16}-\cite{18}. This
approach is based on a diagrammatic technique for nonequilibrium systems and allows
us to obtain separately both interfering and noninterfering parts as
a series over a number of incoherent scatterings.

The generalization of this approach to the case of the presence of a coherent
control field was made in \cite{23} and \cite{24}. In Appendix A we show, as
an example, the double scattering contributions to the differential cross section.
On the basis of these and similar expressions for higher order scattering
contributions, we calculate here the spectral dependence of the enhancement
factor. We consider probe light scattering from ultracold clouds of $\phantom{a}^{87}$Rb
atoms, prepared in a magneto-optical trap after the trapping and repumping
lasers and the quadrupole magnetic field are switched off. All atoms populate
the F=1 hyperfine sublevel of the ground state, while the distribution over Zeeman sublevels
is uniform. The spatial distribution of atoms is assumed to be spherically
symmetric and Gaussian. For the typical conditions of the trap, the Doppler
width is many times smaller than the natural line width of the excited state
and the interatomic distances on average are much larger than the optical
wavelength (dilute medium). This allows us to neglect all effects
associated with atomic motion, and atomic collisions.

Probe radiation is quasi resonant with the $F=1\rightarrow F^{\prime }=1$
transition of the $D_{1}$ line (see Fig. 1) and its polarization can be arbitrary. However, for
definiteness we will consider right or left-handed circularly polarized
light. This light is assumed to be weak; all nonlinear effects connected
with the probe radiation will be neglected. In our calculations this field will
be taken into account only in the first non-vanishing order. Besides the probe
light, the atomic ensemble interacts with a coupling, control field. In this paper
we will consider this field tuned to exact resonance with $F=2\rightarrow
F^{\prime }=1$ transition. Its amplitude is determined by the Rabi frequency
$\Omega _{c}=2|V_{nm^{\prime }}|,$ $V_{nm^{\prime }}$ are the transition
matrix elements for the coupling mode between states $|n\rangle $ and $%
|m^{\prime }\rangle \equiv |F,m^{\prime }\rangle $, which we define with
respect to the $|m^{\prime }\rangle $ $=|{F=2,m^{\prime }=-1>}\rightarrow
|n\rangle =|{F^{\prime }=1,n=0>}$ hyperfine Zeeman transition. Other
transition matrix elements are proportional to $\Omega _{c}$ and algebraic
factors depending on corresponding Clebsch-Gordon coefficients.
\begin{figure}[th]
\begin{center}
{$\scalebox{0.5}{\includegraphics*{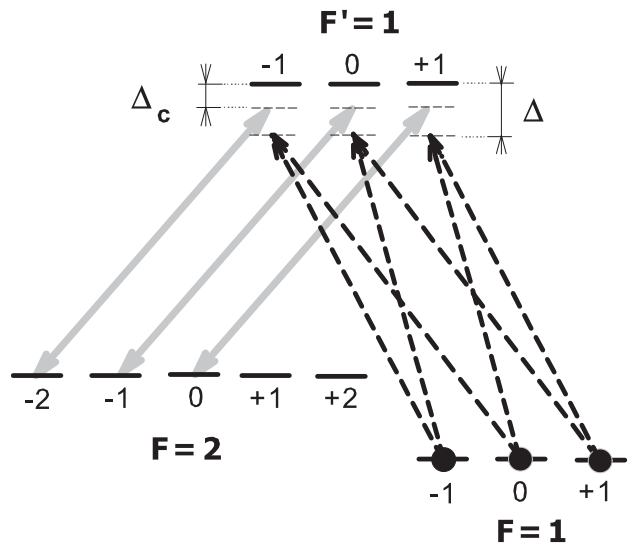}}$ }
\caption{Excitation scheme for observation of the EIT effect in
the system of hyperfine and Zeeman sublevels of the $D_1$-line of
${}^{87}$Rb. The coupling field is applied with right-handed
circular polarization to the $F=2\to F'=1$ transition. The probe
mode is applied to $F=1\to F'=1$ transition and can cause
different excitations depending on its polarization and
propagation direction. }
\end{center}
\par
\label{f1}
\end{figure}
The polarization of the control field is assumed to be right handed and circular. The
detuning of the probe laser frequencies from the corresponding atomic
resonances is assumed to be much less than the hyperfine splitting. This
circumstance, along with the relatively large hyperfine splitting in the excited
state, allows us take into account only one hyperfine sublevel $F' = 1$ of
this state.

In this paper we will not focus our attention on the angular distribution and shape of the CBS
cone, but will instead restrict ourselves by consideration of exact backscattering only.
The ground state hyperfine splitting of ${}^{87} Rb$ is about 6.83 GHz, so Rayleigh
and elastic Raman scattering (associated with the $F=1\rightarrow F^{\prime
}=1\rightarrow F=1$ transition) is well spectrally separated from  inelastically Raman
scattered light $(F=1\rightarrow F^{\prime }=1\rightarrow F=2)$. We will
assume spectral selection under photo detection. The inelastic Raman
component is assumed to be not registered. Depending on the type of
polarization analyzer used for photodetection, four main polarization schemes
can be considered $H_{+}\rightarrow H_{+},\,H_{+}\rightarrow
H_{-},\,H_{-}\rightarrow H_{+},\,H_{-}\rightarrow H_{-}$. Here $H_{\pm }$
represents the helicity of the input and outgoing light. Note that, despite homogeneous
population of the Zeeman sublevels, the susceptibility tensor becomes
essentially anisotropic due to the presence of the coupling field \cite{23},%
\cite{24}. In such a case the enhancement factor depends not only on the relative
polarizations of the input and output light but also on their specific types.

\section{Results}

The results of calculations of the spectrum of the enhancement factor for a weak
control field are shown in Fig. 2a. The calculations were performed for
an atomic cloud with a Gaussian radius equal to $r_{0}=0.5\,cm$; maximal density is $%
n_{0}=3.2\cdot 10^{10}cm^{-3}$. The Rabi frequency of the control field is $\Omega
_{c}=0.5\Gamma $. For a weak control field we observe a not particularly surprising behavior of
the spectrum. Against a background typical for the spectral dependence of CBS
effect we see narrow spectral gap which has its origin in the decreasing of
the optical depth of the cloud caused by the EIT phenomena. Under the EIT effect the
probability of higher order scattering relatively decreases compared with
single scattering and all interference effects which connect with multiple
scattering are reduced. The width of the gap in the spectrum is about the width
of the transparency window determined by the EIT phenomenon. We point out only that
there is a certain difference in the width for different polarization
channels caused by the above-mentioned optical anisotropy of the atomic ensemble.
The situation changes dramatically when the control Rabi frequency becomes comparable to or
larger than the spontaneous decay rate. In Fig. 2b, which is  calculated for $%
\Omega _{c}=3\Gamma $, in addition to the more noticeable anisotropy, an essential
transformation of the spectrum takes place. The range of the structure of this spectrum
connects with the difference in the Autler-Townes splitting for different Zeeman
transitions. The latter is caused by different dipole moments of
the corresponding transitions and consequently with different Rabi frequencies for
them. The maximal value of the enhancement factor for polarization channels with
changing helicity is almost the same as for weak control field but for the
case with preserving helicity we see qualitative modifications. The main one of these is
that, instead of constructive interference for some spectral regions, we
observe destructive interference. In these regions, for channels $H_{-}\rightarrow H_{-}
$ and $H_{+}\rightarrow H_{+}$, the enhancement factor becomes less than unity. That is, in
place of a CBS cone we have a CBS gap, or anticone. In spite of the relatively small value of the
gap it seems physically important because CBS or weak localization itself in
its \textquotedblleft traditional\textquotedblright\ interpretation connects
with time-reversal and always causes enhancement in back scattering.
\begin{figure}[th]
\begin{center}
{$\scalebox{0.5}{\includegraphics*{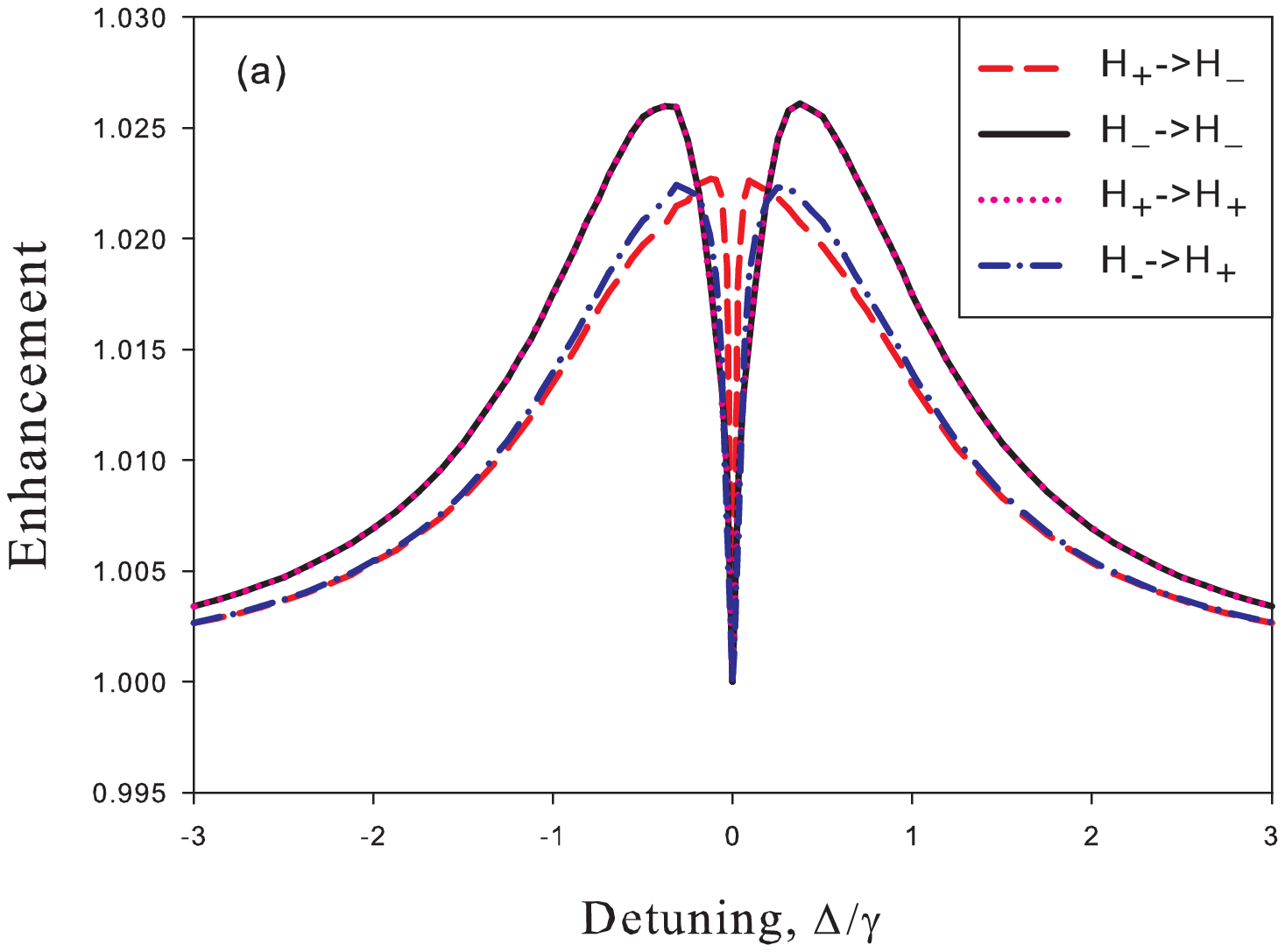}}$ }{$\scalebox{0.5}{%
\includegraphics*{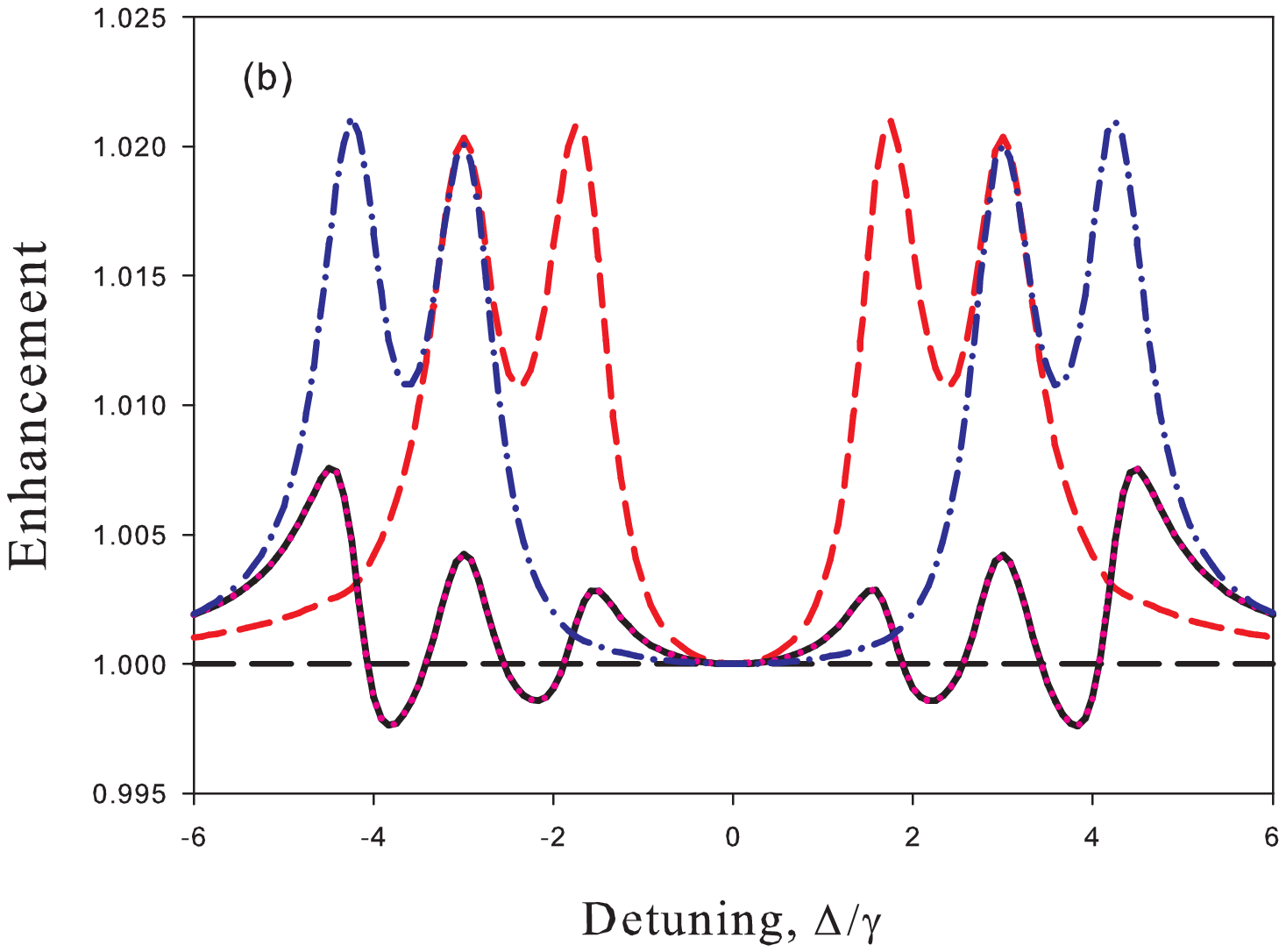}}$ }
\caption{Spectrum of the enhancement factor. (a) $\Omega_{c}=0.5\Gamma $, (b) $\Omega_{c}=3\Gamma $}
\end{center}
\par
\label{fig2}
\end{figure}

A similar effect for scattering of polarized electrons in a solid was shown in
\cite{25}. Following this, antilocalization in electron transport has been widely studied in
a range of physical systems where electrons interact directly with magnetic
impurities and where the spin-orbit interaction is important \cite{26,27}. The
possibility to observe destructive interference in the case of light
scattering from ultracold atomic ensemble was predicted for the first time in \cite{16}.
This effect was explained in \cite{16} by the hyperfine interaction in atoms and by possible
interference of transitions through different hyperfine sublevels of excited
atomic states. In the case considered now, the mechanism of antilocalization  is
different. This is emphasized by Fig. 2b, in which the results are obtained for only one excited state sublevel $F'=1$
without taking into account possible hyperfine interactions.

The fundamental possibility of destructive interference under multiple
scattering connected with the coupling field is illustrated in Fig. 3. Here
we show double backscattering of a positive helicity incoming photon from the probe light beam on a
system consisting of two ${}^{87}$Rb atoms; the exit channel consists of detection
of light also of positive helicity. Here the double scattering is a
combination of the Raleigh-type and Raman-type transitions. In the direct path the scattering
consists of a sequence of Rayleigh-type scattering in the first step and of
Raman-type scattering in the second one. In the reciprocal path, Raman-type
scattering occurs first, and the positive helicity photon undergoes
Rayleigh-type scattering in the second step. There is an
important difference in the transition amplitudes associated with Raleigh
process for these two interfering channels. Indeed, in the direct and
reciprocal path the scattered mode is coupled with different Zeeman
transitions. In the absence of a coupling field these transitions have the
same amplitude and we always have constructive interference. The coupling field
essentially modifies the scattering process and this modification is different
for different Zeeman transitions. For definite frequencies of probe light, the
scattering amplitudes connecting the direct and reciprocal scattering
channels can be comparable in absolute value but can have phases shifted by
an angle close to $\pi$. In this case these two channels suppress each other.
The considered example is not the only one, and appears only in the double scattering channel. There are
some others which lead to constructive interference. In the higher
scattering order the situation is similar. That is why in our results we have
only partial destructive interference and for the considered parameters only a small suppression of
backscattering.
\begin{figure}[th]
\begin{center}
{$\scalebox{0.9}{\includegraphics*{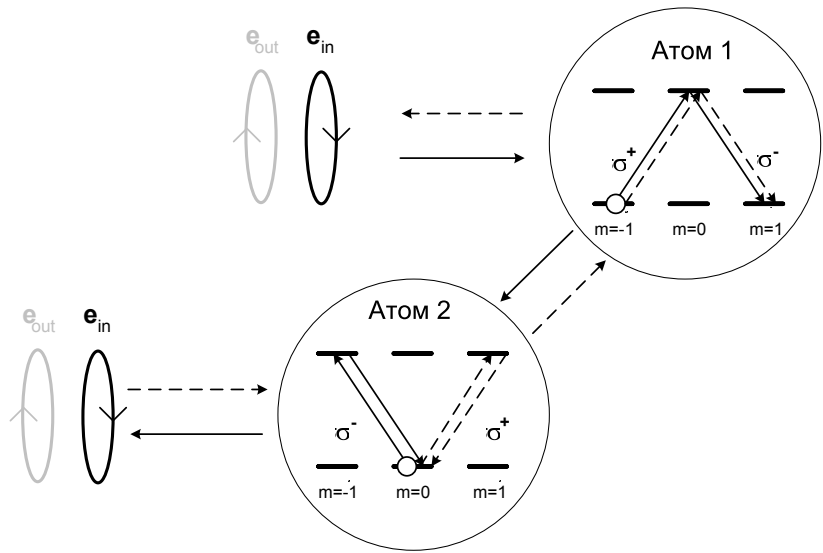}}$ }
\caption{Diagram explaining the antilocalization phenomenon in the example of double scattering in the helicity
preserving scattering channel. Only the interfering transitions initiated by the probe are shown and
the presence of the control mode is specified in Fig. 1. }
\end{center}
\par
\label{f3}
\end{figure}
Note that the antilocalization effect discussed in \cite{16} takes place for
essentially nonresonant light and consequently is difficult for observation,
because it is challenging to prepare clouds with large optical depth for
such radiation. The effect considered here takes place for radiation which is
not far from resonance. It makes this case more promising for experimental
verification. Note also that similar antilocalization phenomena can be
observed in the case when the control field is absent but the probe radiation
is strong enough to make the Autler-Townes splitting noticeable \cite{28}.

Consider now how these peculiarities in the spectrum manifest themselves in time
dependent CBS in the case of a pulsed probe light. Here the most interesting
effects are observed for a weak control field when peculiarities in the
spectrum are sharp. Different spectral components of the input pulse scatter
differently and it is the reason for the essential transformation of the spectrum and
consequently the time profile of the pulse. Scattered light has a deficit of
near resonant photons for which the EIT mechanism works best. The gap in
the spectrum causes light beating effects which are different for scattering of
different orders.  In Fig. 4 we show shapes of pulses which undergo single
(4a) and double (4b) scattering for the $H_{+}\rightarrow H_{-}$ polarization
channel. These graphs are calculated for a cloud with $n_{0}=3.2\cdot
10^{10}cm^{-3}$ and $r_{0}=0.5\,cm$. The input pulse has a Gaussian shape $%
I=I_{0}\exp (-t^{2}/\tau ^{2})$, the length of the pulse is $\tau ^{=}200\Gamma
^{-1}$.
\begin{figure}[th]
\begin{center}
{$\scalebox{0.5}{\includegraphics*{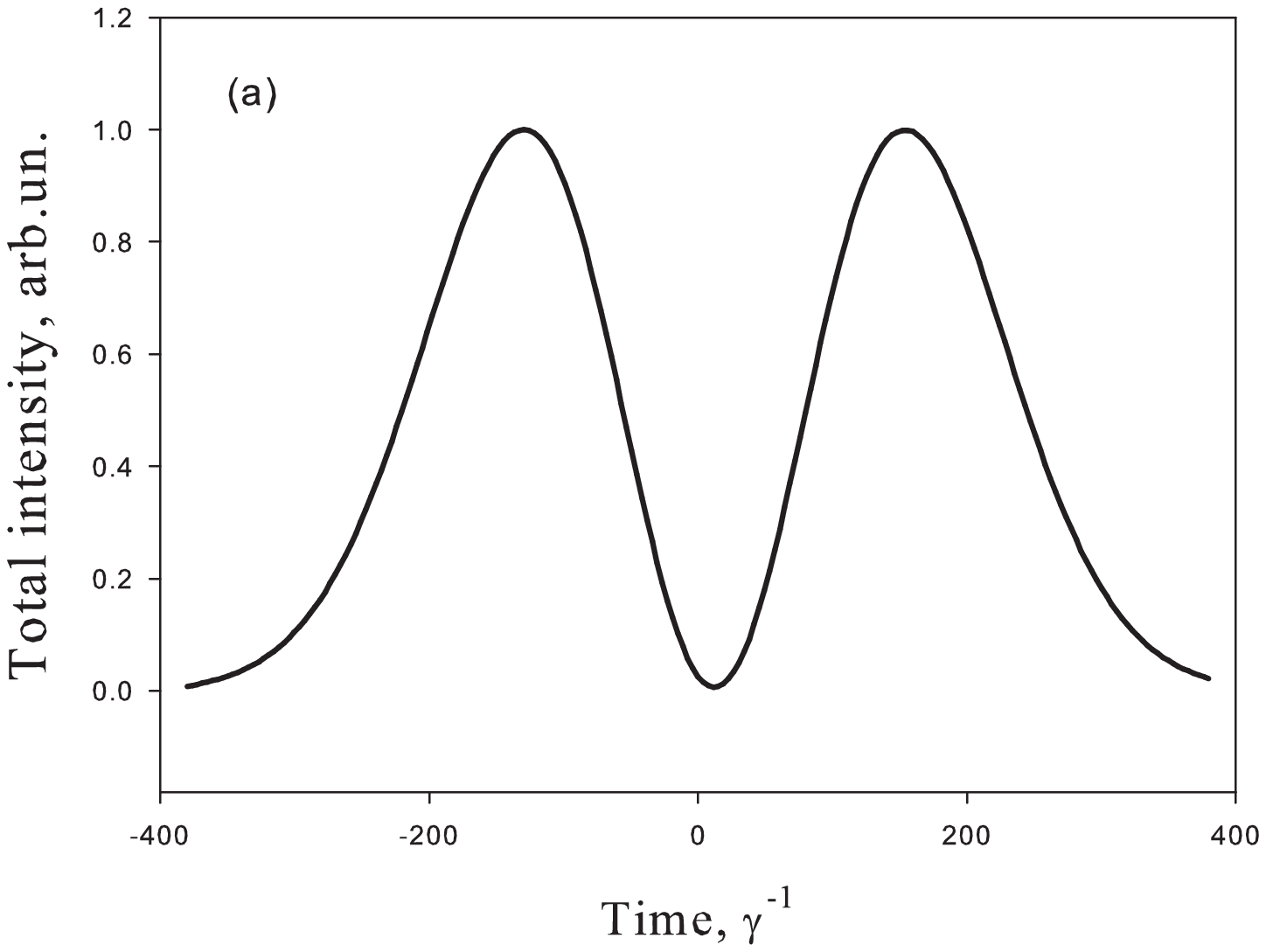}}$ }{$\scalebox{0.5}{%
\includegraphics*{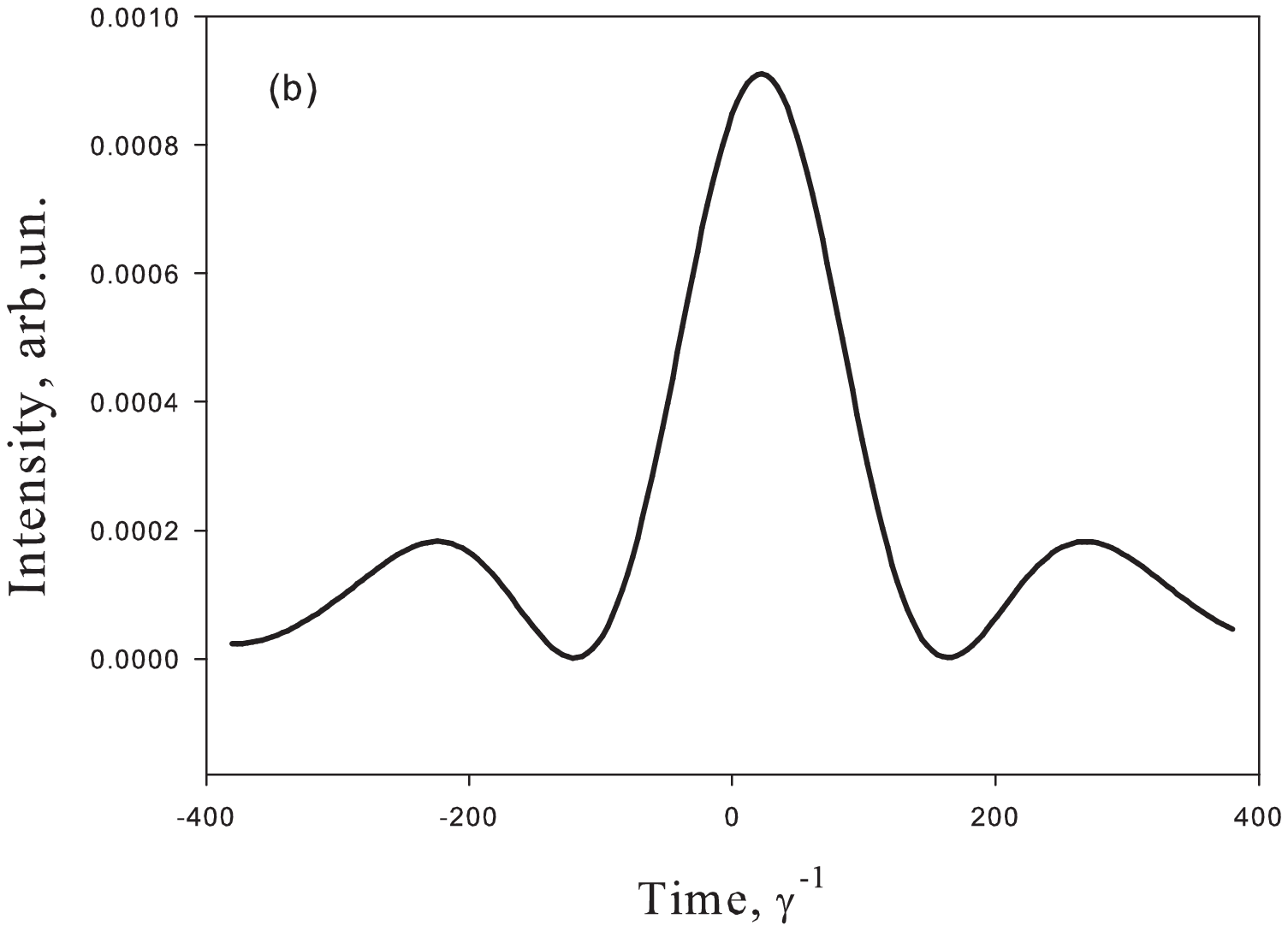}}$ }
\caption{Shapes of the outgoing pulse. (a) single scattering, (b) double scattering}
\end{center}
\par
\label{f4}
\end{figure}

The shapes of single and doubly scattered pulses are essentially
different. Most important is that the intensity peaks for them are separated in time. When single
scattering intensity is maximal double scattering is very small and vice
versa. This effect arises through the relative time scales associated with single scattering in comparison with double scattering, as illustrated in Fig. 5.  For comparison the input pulse shape is also shown.
\begin{figure}[th]
\begin{center}
{$\scalebox{0.5}{\includegraphics*{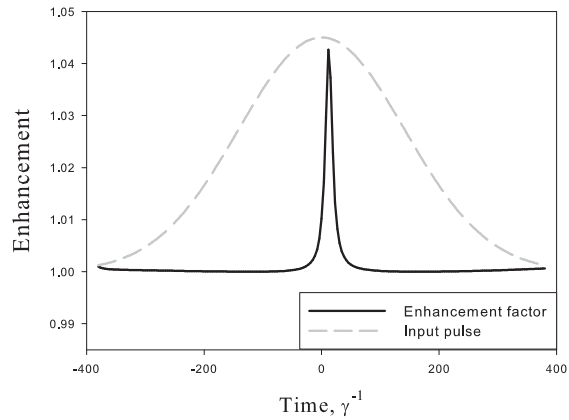}}$ }
\caption{Time dependence of the enhancement factor in the case of pulsed probe radiation. Input pulse shape (in arbitrary
intensity units) is shown in gray}
\end{center}
\par
\label{f5}
\end{figure}

\section*{Acknowledgments}

This work was supported by the Russian Foundation for Basic Research (Grants
08-02-91355 and 10-02-00103).  M.H. acknowledges support from the National
Science Foundation (NSF-PHY-0654226).

\appendix

\section{The cross section of the scattering process}

The scattering process is typically described by the differential cross
section for light scattered into the solid angle $\Omega$. If the system is dilute but optically
dense than the total outcome can be expressed as the sum of the contributions
of the subsequent scattering events, see \cite{12}-\cite{19.0}
\begin{equation}
\frac{d\sigma}{ d\Omega}=\sum_{a=1}^{N} \frac{d\sigma}{ d\Omega}+%
\sum_{a\neq b}\frac{d\sigma_{ab}}{ d\Omega}+%
\sum_{a\neq b\neq c}\frac{d\sigma_{abc}}{d\Omega}+\ldots%
\label{A.1}
\end{equation}
Through macroscopic electrodynamics based self-consistent averaging in the
dilute disordered system, this series is converging rapidly such that any events of
possible recurrent scattering (when the atomic numerator gives $a=c$ etc.) can be
safely ignored. We consider as an example the calculation scheme in the second
order which can be straightforwardly generalized on the higher orders. For the
specific scattering channel in the backward direction the second order contribution
from two randomly selected atoms is given by the sum of two terms
\begin{equation}
\frac{d\sigma _{12}}{d\Omega}=\frac{d\sigma _{12}^{(L)}}{d\Omega} + \frac{d\sigma _{12}^{(C)}}{d\Omega}%
\label{A.2}%
\end{equation}
where the first term is normally called a "ladder" term describing the
contribution of successive double scattering of a photon of frequency
$\omega$, wave vector $\mathbf{k}$, and polarization $\mathbf{e}$ to the
outgoing mode $\omega ^{\prime },\mathbf{k}^{\prime },\mathbf{e}^{\prime }$ on
atom $a=1$ first and atom $b=2$ second. A similar term contributes to the expansion
(\ref{A.1}) for the reciprocal path with $a=2$ and $b=1$. The second "crossed"
term is associated with interference between these scattering events. For the
sake of clarity let us specify the polarization of the incoming photon by
Cartesian unit vector $\mathbf{e}_{\mu}$ with $\mu=x,y$ ($\mathbf{k}\parallel
z$) and of the outgoing photon by $\mathbf{e}'_{\nu}$ with $\nu=x',y'$
($\mathbf{k}'\parallel z'$).

Then scattering in the particular polarization channel from $\mathbf{e}_{\mu}$
to $\mathbf{e}'_{\nu}$ is described by successive double scattering in the
"ladder" contribution and is given by
\begin{eqnarray}
\lefteqn{\frac{d\sigma _{12}^{(L)}}{d\Omega }=\sum\limits_{i,j,k,l}\sum\limits_{\overline{i},\overline{j},\overline{k},
\overline{l}}\frac{\omega _{12}^{4}}{c^{4}}\frac{\omega \omega^{\prime 3}}{c^{4}}\frac{1}{r_{12}^2}}
\notag \\
&&\times X_{\nu i}(\mathbf{\infty },\mathbf{r}_{2},\omega^{\prime })%
\alpha _{ik}^{m_{2}''m_{2}}(\omega _{12}\!-\!\mathbf{k}_{12}\mathbf{v}_{2})%
X_{kl}(\mathbf{r}_{2},\mathbf{r}_{1},\omega _{_{12}})%
\alpha _{lj}^{m_{1}^{\prime \prime }m_{1}}(\omega\!-\!\mathbf{kv}_{1})%
X_{j\mu }(\mathbf{r}_{1}\mathbf{,-\infty ,}\omega )%
\notag \\
&&\times\left[ X_{\nu\overline{i}}(\mathbf{\infty },\mathbf{r}_{2},\omega ^{\prime })%
\alpha _{\overline{i}\,\overline{k}}^{m_{2}''m_{2}}(\omega _{12}\!-\!\mathbf{k}_{ 12}\mathbf{v}_{2})%
X_{\overline{k}\,\overline{l}}(\mathbf{r}_{2},\mathbf{r}_{1},\omega _{12})%
\alpha _{\overline{l}\,\overline{j}}^{m_{1}^{\prime \prime }m_{1}}(\omega \!-\!\mathbf{kv}_{1})%
X_{\overline{j}\mu}(\mathbf{r}_{1}\mathbf{,-\infty ,}\omega )\right]^{*}%
\label{A.3}%
\end{eqnarray}
where the photon's output and virtual intermediate frequency are affected by
both the Raman and residual Doppler effects
\begin{eqnarray}
\omega _{12} &=&\omega \,-\,\omega _{m_{1}''m_{1}}\,+\,(\mathbf{k}%
_{12}-\mathbf{k})\mathbf{v}_{1}  \notag \\
\omega ^{\prime } &=&\omega _{12}\,-\,\omega _{m_{2}''m_{2}}\,+\,(%
\mathbf{k}^{\prime }-\mathbf{k}_{12})\mathbf{v}_{2}%
\label{A.4}%
\end{eqnarray}%
where the following notation is used: $\mathbf{v}_{1}$ and $\mathbf{v}_{2}$ are
the velocities of the first and the second atoms; $\mathbf{k}_{12}$ and
$\mathbf{k} ^{\prime }$ are the photon's wave vectors in the intermediate and
outgoing modes; $r_{12}= |\mathbf{r}_{2}-\mathbf{r}_{1}|$ is the relative
distance between the atoms. For the intermediate mode the wave vector is given
by
$\mathbf{k}_{12}=(\omega_{12}/c)(\mathbf{r}_{2}-\mathbf{r}_{1})/|\mathbf{r}_{2}-\mathbf{r}_{1}|$.

There are several important ingredients contributing to the cross section.
The first is the scattering tensor, which is given by
\begin{eqnarray}
{\alpha}_{lj}^{(m^{\prime \prime }m)}(\Delta)%
=-\sum_{m^{\prime }(n),n}\frac{1}{\hbar }\frac{(d_{l})_{m^{\prime\prime}n}(d_{j})_{nm}}{\Delta+i\Gamma /2}%
\left\{ 1-\frac{|V_{nm^{\prime }}|^{2}}{\Delta+i\Gamma
/2}\,\frac{1}{\Delta _{c}-\Delta+\Sigma _{nm^{\prime }}(\Delta)}%
\right\}%
\label{A.5}%
\end{eqnarray}%
With reference to Fig. 1  the self-energy radiation term $\Sigma
_{nm^{\prime }}(\Delta)=|V_{nm^{\prime }}|^{2}/\left( \Delta+i\Gamma /2\right)$
is expressed by transition matrix elements $V_{nm^{\prime }}$ for the control
mode coupling the non-populated states $|n\rangle$ and $|m^{\prime}\rangle
\equiv |F,m^{\prime }\rangle$ the frequency detunings $\Delta _{c}$ and
$\Delta$ are respectively the offsets of the probe $\omega$ and control $\omega
_{c}$ from the resonance $\Delta=\omega -\omega _{F^{\prime }F}$ ($F=1$) and
$\Delta_c=\omega_c-\omega _{F^{\prime }F}$ ($F=2$); $\Gamma$ is the natural
radiation rate of the excited states; $(d_{j})_{nm}$ represents the dipole transition
moments between the lower $|m\rangle \equiv |F,m\rangle $ and upper $|n\rangle
\equiv |F^{\prime },n\rangle $ states. The scattering tensor determines the
amplitude of the single scattering event for either the elastic or inelastic
channel accompanied by the atomic transition from the state $|m\rangle \equiv
|F,m\rangle $ with $F=1$ to the state $|m^{\prime \prime }\rangle \equiv
|F,m^{\prime \prime }\rangle $ with $F=1$ (Rayleigh channel) or $F=2$
(inelastic Raman channel), see Fig. 1.

Before, between, and after two successive scattering events the photon
propagates in the bulk medium in accordance with the macroscopic Maxwell
theory. The transformation of its amplitude is relevantly described by the
formalism of the Green's propagation function, see \cite{18,23,24}. The matrix
functions $X_{kl}(\mathbf{r}_{2},\mathbf{r}_{1};\omega )$ perform the slowly
varying amplitudes associated with the photon retarded-type Green's function.
They follow the transformation of the light amplitude and polarization between
points $\mathbf{r}_{1}$ and $\mathbf{r}_{2}$. One of these points can approach
infinity for either the input or output scattering channels. The subscript
tensor indices $k$ and $l$ are always associated with the specific frame where
the propagation direction $\mathbf{r}_{2}-\mathbf{r}_{1}$ is associated with
$z$-axis. The components $k$ and $l$ belong to the orthogonal plane $z$ and we
enumerate them by $1$ ($x$-axis) and $2$ ($y$-axis). In this basis the Green's
function slowly varying components can be expressed as follows
\begin{eqnarray}
X_{11}(\mathbf{r}_{2},\mathbf{r}_{1};\omega ) &=&%
e^{i\phi _{0}(\mathbf{r}_{2},\mathbf{r}_{1})}%
\left( \cos \phi (\mathbf{r}_{2},\mathbf{r}_{1})-\sin \phi (\mathbf{r}_{2},\mathbf{r}_{1})n_{x}\right)%
\nonumber \\%
X_{22}(\mathbf{r}_{2},\mathbf{r}_{1};\omega ) &=&%
e^{i\phi _{0}(\mathbf{r}_{2},\mathbf{r}_{1})}%
\left( \cos \phi (\mathbf{r}_{2},\mathbf{r}_{1})+\sin \phi (\mathbf{r}_{2},\mathbf{r}_{1})n_{x}\right)%
\nonumber \\%
X_{12}(\mathbf{r}_{2},\mathbf{r}_{1};\omega ) &=&%
ie^{i\phi _{0}(\mathbf{r}_{2},\mathbf{r}_{1})}\sin \phi (\mathbf{r}_{2},\mathbf{r}_{1})%
\left(n_{y}+in_{z}\right)%
\nonumber\\%
X_{21}(\mathbf{r}_{2},\mathbf{r}_{1};\omega ) &=&%
ie^{i\phi _{0}(\mathbf{r}_{2},\mathbf{r}_{1})}\sin \phi (\mathbf{r}_{2},\mathbf{r}_{1})%
\left(n_{y}-in_{z}\right)%
\label{A.6}%
\end{eqnarray}%
where%
\begin{eqnarray}
\phi _{0}(\mathbf{r}_{2},\mathbf{r}_{1}) =\frac{2\pi \omega }{c}%
\int_{\mathbf{r}_{1}}^{\mathbf{r}_{2}}\chi_0(\mathbf{r},\omega )ds;%
\;\;\;\;\;
\phi (\mathbf{r}_{2},\mathbf{r}_{1}) =\frac{2\pi \omega }{c}%
\int_{\mathbf{r}_{1}}^{\mathbf{r}_{2}}\chi (\mathbf{r},\omega )ds%
\label{A.7}%
\end{eqnarray}%
performs the phase integrals along the path $s$ linking the points
$\mathbf{r}_1$ and $\mathbf{r}_2$ such that $\mathbf{r}=\mathbf{r}(s)$ in the
integrand. The parameters of these integrals are expressed by components of the
dielectric susceptibility tensor of the medium.

Because of the specific symmetry of the problem, see Fig. 1 it
is convenient to define these integrals in the alternative basis set of
circular polarizations. It can be defined by the following expansion
$\mathbf{e}_{0}=\mathbf{e}_{z}$, $\mathbf{e}_{\pm 1}=\mp (\mathbf{e}_{x}\pm
i\mathbf{e}_{y})/\sqrt{2}$ in the Cartesian frame and it requires
co/contravariant notation in writing the tensors products, see \cite{VMK}. The
spectrally dependent parameters $\chi (\mathbf{r},\omega )$, $\chi_{0}(\omega)$
and symbolic vector $\overrightarrow{\chi }(\omega )$ perform the expansion
coefficients of the susceptibility tensor projected on the plane orthogonal to
$\mathbf{r}_{2}-\mathbf{r}_{1}$ in the basis set of the Pauli matrices
$\overrightarrow{\sigma} =(\sigma _{x},\sigma _{y},\sigma _{z})$. This tensor
projection thus can be written
\begin{equation}
\tilde{\chi}_{q}{}^{q^{\prime }}(\mathbf{r},\omega )=\chi_0(\mathbf{r},\omega)%
\delta _{q}{}^{q^{\prime }}+%
\left( \overrightarrow{\chi}(\mathbf{r},\omega)\cdot\overrightarrow{\sigma }\right)_{q}{}^{q^{\prime }}%
\label{A.8}%
\end{equation}%
where $q,q'=\pm 1$. To find the parameters of the phase integrals the components
$\tilde{\chi}_{q}{}^{q^{\prime }}$ in the left side of the expansion
(\ref{A.8}) should be explicitly expressed by the components of the
susceptibility tensor in the laboratory frame $\chi_{q}{}^{q^{\prime }}$. Then
$\chi(\mathbf{r},\omega)$ is the complex length of vector
$\overrightarrow{\chi}(\mathbf{r},\omega)$ and $\overrightarrow{n}(\omega )$ is
its "director"\ , which are given by
\begin{eqnarray}
\chi^2(\mathbf{r},\omega)&=&\chi_x^2(\mathbf{r},\omega)+\chi_y^2(\mathbf{r},\omega)+\chi_z^2(\mathbf{r},\omega)%
\nonumber\\%
\overrightarrow{n}&=&\overrightarrow{n}(\omega)=%
\overrightarrow{\chi }(\mathbf{r},\omega)/\chi (\mathbf{r},\omega )%
\label{A.9}%
\end{eqnarray}
We additionally assume that the atomic polarization is homogeneous along the
atomic sample such that the "director" $\overrightarrow{n}(\omega )$ is constant
along the path in the phase integrals representation of the Green's function
(\ref{A.6}).

The local reference frame is linked with the laboratory frame via a rotational
transformation characterized by Euler angles $\alpha ,\beta ,\gamma $. Then the
susceptibility tensor components projected onto the plane orthogonal to the light
ray direction and defined in the basis of circular polarizations, are given by
\begin{eqnarray}
\tilde{\chi}_{+1}{}^{+1}({\scriptstyle\ldots }) &=&\frac{(1+\cos \beta )^{2}%
}{4}\chi _{+1}^{+1}({\scriptstyle\ldots })+\frac{(1-\cos \beta )^{2}}{4}\chi_{-1}^{-1}({\scriptstyle\ldots })%
+\frac{\sin ^{2}\beta }{2}\chi _{0}^{0}({\scriptstyle\ldots }),  \notag \\
\tilde{\chi}_{-1}{}^{-1}({\scriptstyle\ldots }) &=&\frac{(1-\cos \beta )^{2}%
}{4}\chi _{+1}^{+1}({\scriptstyle\ldots })+\frac{(1+\cos \beta )^{2}}{4}\chi
_{-1}^{-1}({\scriptstyle\ldots })%
+\frac{\sin ^{2}\beta }{2}\chi _{0}^{0}({\scriptstyle\ldots }),  \notag \\
\tilde{\chi}_{-1}{}^{+1}({\scriptstyle\ldots }) &=&\frac{1}{4}\mathrm{e}^{2%
\mathrm{i}\gamma }\sin ^{2}\beta \left[ \chi _{+1}^{+1}({\scriptstyle\ldots }%
)+\chi _{-1}^{-1}({\scriptstyle\ldots })-2\chi _{0}^{0}({\scriptstyle\ldots})\right]%
\notag \\%
\tilde{\chi}_{+1}{}^{-1}({\scriptstyle\ldots }) &=&\frac{1}{4}\mathrm{e}^{-2%
\mathrm{i}\gamma }\sin ^{2}\beta \left[ \chi _{+1}^{+1}({\scriptstyle\ldots }%
)+\chi _{-1}^{-1}({\scriptstyle\ldots })-2\chi _{0}^{0}({\scriptstyle\ldots})\right]%
\label{A.10}%
\end{eqnarray}
where in the right side the tensor components
$\chi_{q}^{q^{\prime}}({\scriptstyle\ldots })$ are associated with the
laboratory frame. In the basis of complex polarizations
$\mathbf{e}_{0},\mathbf{e}_{\pm 1}$  this tensor has a
symmetric diagonal structure and its components are given by
\begin{eqnarray}
\chi _{q}^{q^{\prime }}(\mathbf{r},\Delta) =-\delta _{q}^{q^{\prime}}%
\frac{n_0(\mathbf{r})}{2F+1}\!\!\sum_{n(m),m^{\prime }(m),m}\!\!%
\frac{1}{\hbar }\frac{|(\mathbf{d}\mathbf{e}_{q}^{\ast })_{nm}|^{2}}{\Delta+i\Gamma /2}%
\left\{ 1-\frac{|V_{nm^{\prime }}|^{2}}{\Delta+i\Gamma /2}\,%
\frac{1}{\Delta _{c}-\Delta+\Sigma _{nm^{\prime}}(\Delta)}\right\}%
\label{A.11}
\end{eqnarray}%
where $n_{0}(\mathbf{r})$ is the local density of atoms and the states
$|n\rangle$ and $|m'\rangle$ in the sum are unique for each $|m\rangle$ such
that $n=n(m)$ and $m'=m'(m)$.

The "crossed" term contribution in the differential cross section (\ref{A.2})
is expressed similarly to the "ladder" term
\begin{eqnarray}
\lefteqn{\frac{d\sigma _{12}^{(C)}}{d\Omega }=\sum\limits_{i,j,k,l}\sum\limits_{\overline{i},\overline{j},\overline{k},
\overline{l}}\frac{\omega_{12}^{2}}{c^{2}}\frac{\omega_{21}^{2}}{c^{2}}%
\frac{\omega \omega^{\prime 3}}{c^{4}}\frac{1}{r_{12}^2}%
\exp\left[i(%
\mathbf{k+k}^{\prime }\mathbf{)(r}_{1}-\mathbf{r}_{2})%
+i\frac{(\omega_{12}-\omega _{21})}{c}r_{12}\right]}
\notag \\%
&&\times X_{\nu i}(\mathbf{\infty },\mathbf{r}_{2},\omega^{\prime })%
\alpha _{ik}^{m_{2}''m_{2}}(\omega _{12}\!-\!\mathbf{k}_{12}\mathbf{v}_{2})%
X_{kl}(\mathbf{r}_{2},\mathbf{r}_{1},\omega _{_{12}})%
\alpha _{lj}^{m_{1}^{\prime \prime }m_{1}}(\omega\!-\!\mathbf{kv}_{1})%
X_{j\mu }(\mathbf{r}_{1}\mathbf{,-\infty ,}\omega )%
\notag \\
&&\times\left[ X_{\nu\overline{i}}(\mathbf{\infty },\mathbf{r}_{1},\omega ^{\prime })%
\alpha _{\overline{i}\,\overline{k}}^{m_{1}''m_{1}}(\omega _{21}\!-\!\mathbf{k}_{21}\mathbf{v}_{1})%
X_{\overline{k}\,\overline{l}}(\mathbf{r}_{1},\mathbf{r}_{2},\omega _{21})%
\alpha _{\overline{l}\,\overline{j}}^{m_{2}''m_{2}}(\omega \!-\!\mathbf{kv}_{2})%
X_{\overline{j}\mu}(\mathbf{r}_{2}\mathbf{,-\infty ,}\omega )\right]^{*}%
\label{A.12}%
\end{eqnarray}
where for the reciprocal path the frequency $\omega _{21}$ given by
\begin{equation}
\omega _{21}=\omega -\omega _{m_{2}''m_{2}}+(\mathbf{k}_{21}-\mathbf{k})\mathbf{v}_{2}%
\label{A.13}%
\end{equation}
is different from $\omega_{12}$ but $\omega'$ for the scattering in the
backward direction is the same.

In the macroscopic and disordered atomic system the "crossed" term (\ref{A.12})
contributes mainly in the narrow solid angle near backward direction
($\mathbf{k}^{\prime }\approx -\mathbf{k}$). For any other directions this term
makes negligible contribution. In higher orders of multiple scattering the
"ladder" and interference "crossed" parts have the similar structure. All these
term can be subsequently (moving from lower to higher orders) involved into the
Monte Carlo simulation scheme. The cross section should be averaged with the atomic
density matrix over the initial states $m_1$, $m_2$,... and over the atomic velocity
distribution for $\mathbf{v}_1$, $\mathbf{v}_2$ ... This averaging is
equivalent to the smoothed sum in the original expansion of the cross section
given by Eq.(\ref{A.1}).

\end{document}